\documentclass[11pt,twoside]{article}

\usepackage{asp2006}
\usepackage{epsf}
\usepackage{lscape}
\usepackage{graphicx}

\begin{document}
\title{Reconnection Diffusion and Star Formation Processes}   
\author{A. Lazarian\altaffilmark{1}, R. Santos-Lima\altaffilmark{1,2}, and E. de Gouveia Dal Pino\altaffilmark{2}}   
\affil{$^{1}$Department of Astronomy, University of Wisconsin-Madison, USA (lazarian@astro.wisc.edu),\\
$^{2}$ Instituto de Astronomia, Geofisica e Ciencias Atmosfericas, S\~ao Paulo, Brazil}

\begin{abstract} 
The diffusion of astrophysical magnetic fields in conducting fluids in the presence of turbulence depends on whether magnetic fields can change their topology or reconnect in highly conducting media. Recent progress in understanding fast magnetic reconnection in the presence of turbulence is reassuring that the magnetic field behavior in computer simulations and turbulent astrophysical environments is similar, as far as the magnetic reconnection is concerned. This makes it meaningful to perform MHD simulations of turbulent flows in order to understand the diffusion of magnetic field in astrophysical environments. These simulations support the concept of reconnection diffusion, which describes the ability of magnetic fields to get removed from magnetized clouds and cores in the process of star formation. 
\end{abstract}


\section{Magnetic Reconnection}   

\begin{figure}[!t]
 \begin{center}
\includegraphics[width=.6 \columnwidth]{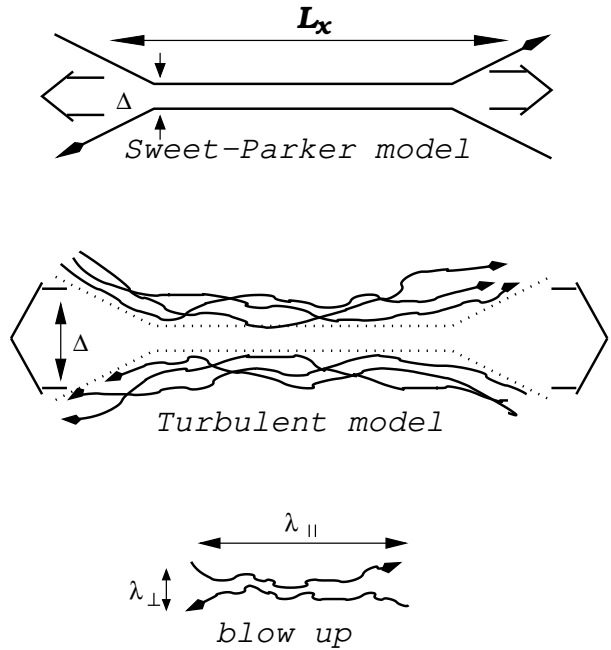}
\caption{{\it Upper plot}:
Sweet-Parker model of reconnection. The outflow
is limited by a thin slot $\Delta$, which is determined by Ohmic
diffusivity. The other scale is an astrophysical scale $L\gg \Delta$.
{\it Middle plot}: Reconnection of weakly stochastic magnetic field according to
LV99. The model that accounts for the stochasticity
of magnetic field lines. The outflow is limited by the diffusion of
magnetic field lines, which depends on field line stochasticity.
{\it Low plot}: An individual small scale reconnection region. The
reconnection over small patches of magnetic field determines the local
reconnection rate. The global reconnection rate is substantially larger
as many independent patches come together. From Lazarian et al. 2004.}
\label{recon1}
 \end{center}
\end{figure}

Magnetic reconnection describes the ability of magnetic field to change its topology. The famous Sweet-Parker model of reconnection (Sweet 1958, Parker 1957) (see Figure~\ref{recon1}, upper panel) produces reconnection rates which are smaller than the Alfv\'en velocity by a square root of the
Lundquist number, i.e. by $S^{-1/2}\equiv (LV_A/\eta)^{-1/2}$, where $L$ in this case is the length of the current sheet. Thus this scheme produces reconnection at a rate which is negligible for most of astrophysical circumstances. If the Sweet-Parker were proven to be the only possible model of reconnection, it would have been possible to show that MHD numerical simulations do not have anything to do with real astrophysical fluids. Fortunately, faster schemes of reconnection are available.

The first model of fast reconnection proposed by Petschek (1964) assumed that magnetic fluxes get into contact not along the astrophysically large scales of $L$, but instead over a scale comparable to the resistive thickness $\delta$, forming a
distinct X-point, where magnetic field lines of the interacting fluxes converge at a sharp point to the reconnection spot. The stability of such a reconnection geometry in astrophysical situations is an open issue. At least for uniform resistivity, this configuration was proven to be unstable and to revert to a Sweet-Parker configuration (see Biskamp 1986, Uzdensky \& Kulsrud 2000).

Recent years have been marked by the progress in understanding some of the key processes of reconnection in astrophysical plasmas. In particular, a substantial progress has been obtained by considering reconnection in the presence of the Hall-effect (see Shay et al. 1998).
The condition for which the Hall-MHD term becomes important for the reconnection is that the ion skin depth $\delta_{ion}$ becomes comparable with the Sweet-Parker diffusion scale $\delta_{SP}$. The ion skin depth is a microscopic characteristic and it can be viewed as the gyroradius of an ion moving at the Alfv\'en speed, i.e. $\delta_{ion}=V_A/\omega_{ci}$, where $\omega_{ci}$ is the cyclotron frequency of an ion. For the parameters of the interstellar medium (see Table~1 in Draine \& Lazarian 1998), the reconnection is collisional (see further discussion in Yamada 2006).

A radically different model of reconnection was proposed in Lazarian \& Vishniac (1999, henceforth LV99). The middle and bottom panels of Figure~\ref{recon1} illustrate the key components of LV99 model\footnote{The cartoon in Figure~\ref{recon1} is an idealization of the reconnection process as the actual reconnection region also includes reconnected open loops of magnetic field moving oppositely to each other. Nevertheless, the cartoon properly reflects the role of the 3-dimensionality of the reconnection process, the importance of small-scale reconnection events, and the increase of the outflow region compared to the Sweet-Parker scheme.}. The reconnection events happen on small scales $\lambda_{\|}$ where magnetic field lines get into contact. As the number of independent reconnection events that take place simultaneously is $L/\lambda_{\|}\gg 1$ the resulting reconnection speed is not limited by the speed of individual events on the scale $\lambda_{\|}$. Instead, the constraint on the reconnection speed comes from the thickness of the outflow reconnection region $\Delta$, which is determined by the magnetic field wandering in a turbulent fluid. The model is intrinsically three dimensional as both field wandering and simultaneous entry of many independent field patches, as shown in Figure~\ref{recon1}, are 3D effects. The magnetic reconnection speed becomes comparable with $V_A$ when the scale of magnetic field wandering $\Delta$ becomes comparable with $L$.

The LV99 model was successfully tested in Kowal et al. (2009) and we shall use this model to justify the concept of ``reconnection diffusion'' below. The advantage of LV99 model is that it is applicable to both collisional and collisionless plasmas.

\section{Reconnection diffusion}
\label{sec:model}

Magnetic reconnection was appealed in Lazarian (2005) as a way of removing magnetic flux from gravitating clouds, e.g. from star forming clouds. The paper referred to the reconnection model in LV99 and Lazarian, Vishniac \& Cho (2004) for the justification of the concept of fast magnetic reconnection in the presence of turbulence. The advantage of the scheme in Lazarian (2005) was that the removal of magnetic flux can be accomplished both in partially and fully ionized plasma with the robust removal, which were only marginally dependent on the ionization state of the gas\footnote{The rates were predicted to depend on the reconnection rate, which according to Lazarian et al. (2004) marginally depends on the ionization degree of the gas.}. The idea of ``reconnection diffusion'' deserves more attention now as the basic idea of the LV99 model has been confirmed.

Compared to the concept of ``ambipolar diffusion'', which is usually appealed in the star formation community as a way of removing magnetic flux, the ``reconnection diffusion'' has several advantages. First of all, it does not sensitively depend on the ionization rate and therefore can remove magnetic flux from the material with high degree of ionization. The requirement for the ``reconnection diffusion'' to  happen is the presence of turbulence, which is not restrictive for most astrophysical environments, including molecular clouds. 

We should also state that unlike transient redistribution of magnetic field and matter discussed in Cho \& Lazarian (2002, 2003) and Passot \& Vazquez-Semadeni (2003) in terms of fast, slow modes, the ``reconnection diffusion" provides a permanent separation of magnetic field and fluids. Thus we expect the latter 
concept to be applicable to a variety of situations, including accretion discs, molecular cloud cores etc.

\section{Numerical simulations of ``reconnection diffusion''}

To test the concept of ``reconnection diffusion'' we performed resistive one fluid MHD simulations. The magnetic Reynolds   and Lundquist numbers of typical astrophysical fluids are much larger than those that can be achieved in computer simulations. Therefore it is the fast magnetic reconnection in the presence of turbulence predicted LV99 (see Figure~1) and confirmed numerically in Kowal et al. (2009), that provides the justification of our relating resistive MHD simulations in the presence of turbulence and astrophysics. Indeed, the fast reconnection makes the process of magnetic turbulent diffusion and mixing independent of the actual resistivity and thus from the Lundquist number of the astrophysical flow\footnote{The advantage of the LV99 model of reconnection is that the reconnection speed does not depend on the plasma being collisional or collisionless or any other plasma properties. The reconnection rate depends by the width of the outflow region $\Delta$ (see Figure 1), which is determined by magnetic field wandering. A set of simulations in Kowal et al. (2009) where anomalous resistivity was used to emulate plasma effects confirmed that in the presence of turbulence such effects do not change the reconnection rate.}. To make sure that the diffusion that we observe is not due to purely Ohmic effects, we performed a set of simulations without turbulence which clearly demonstrated that the Ohmic diffusion is negligible on the time scales of our simulations.

 A more extended discussion of our simulations can be found in Santos-Lima et al. (2010). First of all, we have performed simulations of ``reconnection diffusion'' without gravitational field. We found that in the presence of turbulence magnetic field efficiently diffuses in the volume changing the flux to mass ratio for the initial magnetic inhomogeneities. Our results resembled the outcome of two fluid 2.5D simulations in Heitch et al. (2004) with the important difference that their simulations were two fluid ones and thus the fast diffusion there was interpreted as the result of "turbulent ambipolar diffusion". As our code does not have ambipolar diffusion, we interpret the similarity of our and Heitch et al. (2004) results as the evidence that in turbulent fluids ambipolar  diffusion may not play a critical role for removing magnetic flux.

In order to get an insight into the magnetic field diffusion in a turbulent fluid immersed in a gravitational potential, we have performed experiments in the presence of a gravitational potential with cylindric symmetry  $\Psi$, given in cylindrical coordinates $(R, \phi, z)$ by:
\begin{equation}
\Psi (R \leq R_{max}) =  - \frac{A}{R + R_{*}}
\label{eq:pot1}
\end{equation}
\begin{equation}
\Psi (R > R_{max}) =  - \frac{A}{R_{max} + R_{*}}, \label{eq:pot2}
\end{equation}
where $R=0$ is the center of the $(x,y)$-plane, and we fixed $R_{*} = 0.1 L / 2$ and $R_{max} = 0.45 L / 2$. Axially symmetric magnetic field flux is initially concentrated the at center of the gravitational potential, where the density is also at its maximum (see Figure~2). 

These simulations explore the diffusion of magnetic flux in idealized clouds or accretion disks. Focusing on identifying the effects of the ``reconnection diffusion'' we do not attempt to present other features of our systems, e.g. partial ionization of the gas. In other words, our simulations address a question of whether magnetic flux can be removed from a turbulent cloud without any effect of ambipolar diffusion. A positive answer to this question solves many of the outstanding issues of the mordern star formation paradigm.  

\begin{figure*}[!hbt]
 \begin{center}
 \includegraphics[width=1.0 \textwidth]{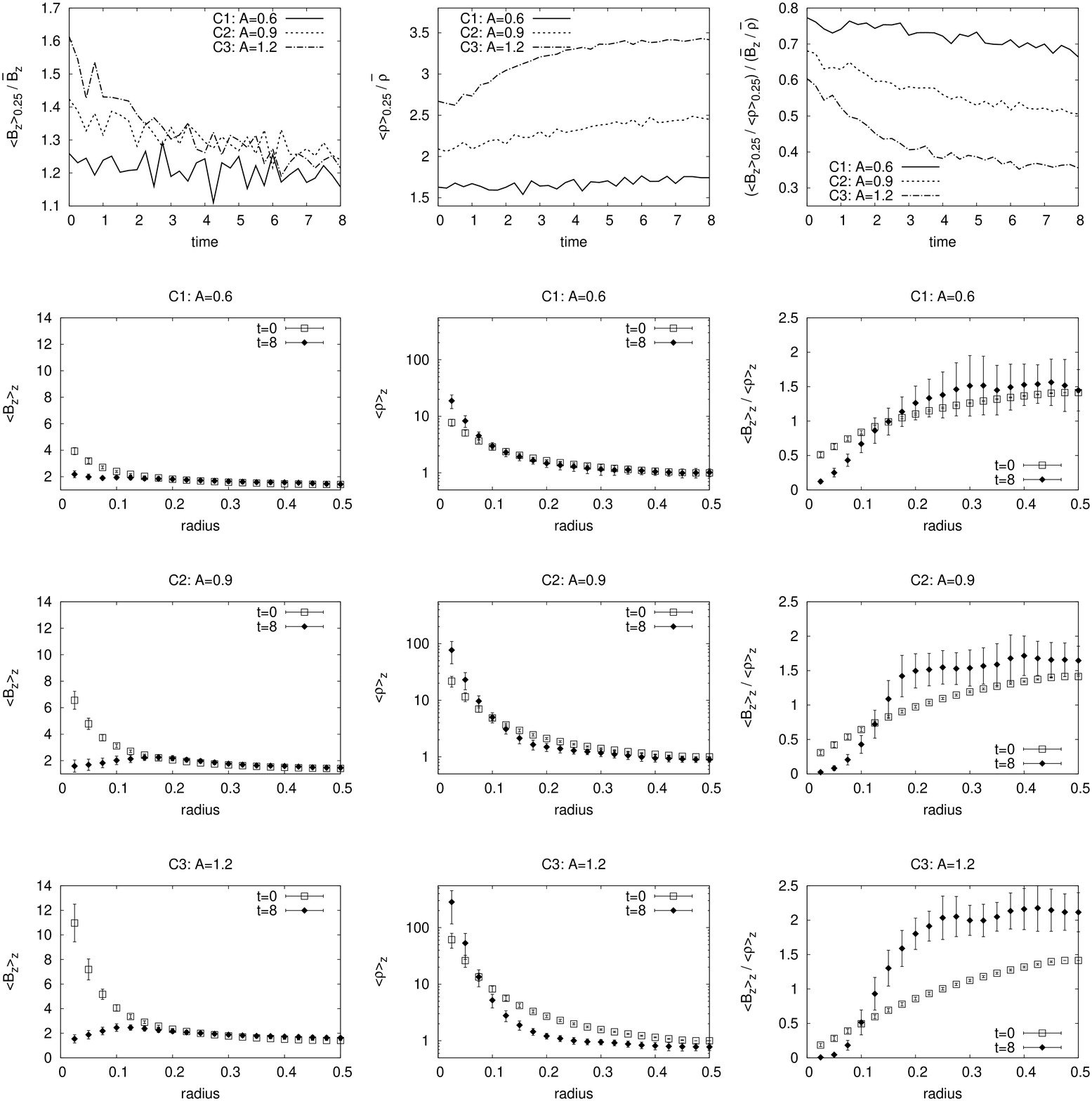}
 \caption{The top row shows the time evolution of $\left\langle B_{z} \right\rangle _{0.25} / \bar{B_{z}}$ (\textit{left}), $\left\langle \rho \right\rangle _{0.25} / \bar{\rho}$ (\textit{middle}), and $(\left\langle B_{z} \right\rangle _{0.25} / \left\langle \rho \right\rangle _{0.25}) / (\bar{\rho} / \bar{B_{z}})$ (\textit{right}). The other rows show the radial profile of $\left\langle B_{z} \right\rangle _{z}$ (\textit{left}), $\left\langle \rho \right\rangle _{z}$ (\textit{middle}), and $\left\langle B_{z} \right\rangle _{z} / \left\langle \rho \right\rangle _{z}$ (\textit{right}) for the different values of $A$ in $t=0$ and $t=8$. Error bars show the standard deviation. All models have initial $\beta=1.0$. Value $A$ is defined by Eq. (1) and (2). }
 \label{fig:grav2}
 \end{center}
\end{figure*}

The result of our simulations are presented in Figure 2, which testifies the decrease of the flux to mass ratio in the system with time. As in the presence of the turbulence both magnetic field and density fluctuates with time, we show the data averaged over the cylinder with the main axis in the direction of the initial magnetic field centered in the maximum of the gravitational potential and with the radius of 0.25 of the computational box. This is further denoted by $\langle \rangle_{0.25}$. 
Top row of Figure \ref{fig:grav2} shows the evolution of $\left\langle B_{z} \right\rangle _{0.25}$ (\textit{left}), $\left\langle \rho \right\rangle _{0.25}$ (\textit{middle}) and $\left\langle B_{z} \right\rangle _{0.25} / \left\langle \rho \right\rangle _{0.25}$ (\textit{right}), normalized by the respective characteristic values inside the box ($\bar{B_{z}}$, $\bar{\rho}$ and $\bar{B_{z}}/\bar{\rho}$), for the models C1, C2, and C3 ($\beta=1$).
We compare the evolution of these quantities for different strengths of gravity $A$, maintaining the other parameters
identical. The central magnetic flux reduces faster the higher the value of $A$. The flux-to-mass ratio has similar
behavior. The other plots of Figure \ref{fig:grav2} show the profile of the quantities
$\left\langle B_{z} \right\rangle _{z}$ (\textit{upper panels}),
$\left\langle \rho \right\rangle_{z} $ (\textit{middle panels}),
and $\left\langle B_{z} \right\rangle _{z} / \left\langle \rho \right\rangle _{z}$ (\textit{bottom panels})
along the radius $R$, each column corresponding to a different value of $A$ both for $t=0$
(in magneto-hydrostatic equilibrium and constant $\beta$) and for $t=8$.
We see  the deepest decay of the magnetic flux toward the central region for the highest value of $A$ at $t=8$.

All in all, we clearly see that turbulence substantially influences the quasi-static evolution of magnetized gas in the gravitational potential. The system in the presence of turbulence relaxes fast to its minimum potential energy state. This explains the change of the flux-to-mass ratio, which for years was thought to a problem that can be dealt only via invoking ambipolar diffusion.

In addition, our simulations performed for the models where the original state was a non-equilibrium one showed that the concept of "reconnection diffusion" carries over to the situations when the magnetized matter evolves on the free-fall time scales in the presence of gravity. While in the absence of turbulence the mass to flux ratio was preserved, this ratio was changing in the presence of turbulence. We note, that while in numerical simulations we may have a laminar collapse, this situation is unlikely for realistic high Reynolds number astrophysical flows.  

In a separate set of simulations we found that in terms of the removal of the magnetic field from quasi-static clouds, the effect of reconnection diffusion is similar to the effect of diffusion induced Ohmic effects. Indeed, our simulations in the absence of turbulence but with strongly enhanced resistivity showed the time evolution of magnetic flux which is similar to our reconnection diffusion runs. In the absence of turbulence and with normal resistivity of our simulations the magnetic diffusion was demonstrated to be negligible.  
  
\section{Discussion}
"Reconnection diffusion" is a process which is expected to happen in the presence of turbulent reconnection predicted by LV99 model. As magnetic turbulence is ubiquitous in astrophysical fluids, we expect "reconnection diffusion" to be ubiquitous as well. The process may remove magnetic flux from star forming clouds and clumps on times much faster than it is allowed by the traditional ambipolar diffusion. 

We note that "reconnection diffusion" should be distinguished from the concept of "turbulent magnetic diffusivity" which is frequently discussed in the framework of kinematic, i.e. without backreaction, dynamo (see Parker 1979). The former concept is based on the tested idea of fast reconnection of strong magnetic fields in the presence of weak turbulence, while the latter concept assumes the efficient diffusion turbulent of magnetic field without taking into account its backreaction, as if the magnetic field were a passive scalar. A usual ``justification'' of the ``turbulent magnetic diffusivity'' is that turbulence mixes magnetic field opposite polarity on the very small scales, which is the process prohibited on the energetic grounds for any dynamically significant field. Therefore we claim that the "turbulent magnetic diffusivity" is an erroneous idea, while the "reconnection diffusion" is a well founded concept.

"Reconnection diffusion" has direct relation to the problem of dissipation in accretion discs  discussed in Shu et al. (2006, henceforth SX06). They have found that the dissipation there should be about four orders of magnitude larger than the Ohmic dissipation in order to solve the magnetic flux problem in these systems. They then appealed to a  hyper-resistivity concept in order to explain the higher dissipation of magnetic field in a turbulent environment.

We feel, however, that the hyper-resistivity idea is poorly justified (see criticism of it in Lazarian et al. 2004). At the same time, fast 3D reconnection can provide the magnetic diffusivity that is required for removal of the magnetic flux. This is what, in fact, was demonstrated in the present set of numerical simulations. It is worth mentioning that, unlike the actual Ohmic diffusivity, magnetic diffusivity mediated by fast reconnection does not transfer the magnetic energy {\it directly} into heat. The lion share of the energy is being released in the form of kinetic energy, driving turbulence. The annihilation of the magnetic field happens in LV99 model, as in any model of fast reconnection, over a small fraction of the volume. This fraction goes to zero as the resistivity goes to zero. Magnetic turbulence induced by reconnection eventually dissipates energy, resulting in the medium heating. If the system is initially laminar, this potentially can result in flares of reconnection and the corresponding diffusivity.

Similar to SX06, we expect to observe the heating of the media. Indeed, although we do not expect to have Ohmic heating, the kinetic energy released due to magnetic reconnection is dissipated locally and therefore we expect to observe heating in the medium.
Our setup for gravity can be seen as a toy model representing the situation in SX06. In the broad sense, our work confirms that a process of magnetic field diffusion that does not rely on ambipolar diffusion is efficient.

We showed that the higher the strength of the gravitational force, the lower is the flux-to-mass ratio in the central region (compared with the mean value in the computational domain). This could be understood in terms of the potential energy of the system. When the potential is higher, more it is energetically  favorable is to pile up of matter near the center of gravity, decreasing the total potential energy of the system. When the turbulence is increased, there is an initial trend to remove more magnetic flux from the center (and consequently more inflow of matter into the center), but for the highest value of the turbulent velocity in our experiments, there is a trend to remove material (together with magnetic flux) from the center, reducing the role of the gravity, due the fact that the gravitational energy became small compared to the kinetic energy of the system. Our results also showed that when the gas is less magnetized (higher $\beta$, or higher values of the Alfv\'enic Mach number $M_{A}$), the turbulent diffusion of magnetic flux is more efficient, but the central flux-to-mass ratio relative to external regions is  smaller for more magnetized models (low $\beta$), compared to less magnetized models. That is, the contrast $B / \rho$ between the inner and outer radius is higher for lower $\beta$ (or $M_{A}$).

If the turbulent diffusivity of magnetic field may explain the results in SX06, one may wonder whether one can remove magnetic field this way not only from the class of systems studied by SX06, but also from less dense systems.
For instance, it is frequently assumed that only ambipolar diffusion is important for the evolution of subcritical magnetized clouds Tassis \& Mouschovias (2005). Our study indicates that this conclusion may require modification in the presence of turbulence. 

While the concept of "reconnection diffusion" describes the diffusion of magnetic field in turbulent media, it is also closely connected to other important astrophysical concepts, i.e. turbulent advection of heat in the presence of turbulence (see Cho et al. 2003). If reconnection were slow, the mixing motions required by the turbulent advection would be difficult to explain. Thus, the fast diffusion of magnetic field induced by turbulence and the turbulent advection of heat in magnetized plasmas are interconnected.

\acknowledgements 
The research of AL is supported by the Center for Magnetic Self-Organization in Laboratory and Astrophysical Plasmas and NSF Grant AST-0808118. 



\begin{thebibliography}
\bibitem[Biskamp(1986)]{biskamp1986} Biskamp, D.\ 1986, Physics of
Fluids, 29, 1520
\bibitem[Cho 
\& Lazarian(2002)]{2002PhRvL..88x5001C} Cho, J., \& Lazarian, A.\ 2002, Physical Review Letters, 88, 245001 
\bibitem[Cho et al.(2003)]{cho2003} Cho, J., Lazarian, A., Honein, A., Knaepen, B., Kassinos, S., \& Moin, P.\ 2003, \apjl, 589, L77
\bibitem[Cho, Lazarian \& Vishniac(2003)]{cho_lazarian_vishniac2003} Cho, J., Lazarian, A., \& Vishniac, E.~T.\ 2003, \apj, 595, 812
\bibitem[Draine\& Lazarian(1998)]{draine1998} Draine, B.~T., \& Lazarian, A.\ 1998, \apj, 508, 157
\bibitem{36a} Drake, J., Swisdak, M., Che, H. \& Shay, M. 2006b, Nature, 443, 05116
\bibitem[Goldreich \& Sridhar(1995)]{goldreich95} Goldreich, P. \& Sridhar, S. \ 1995, \apj 438, 763
\bibitem[Heitsch et al.(2004)]{heitsch2004} Heitsch, F., Zweibel,
E.~G., Slyz, A.~D., \& Devriendt, J.~E.~G.\ 2004, \apj, 603, 165
\bibitem[Kowal et al.(2009)]{kowal2009} Kowal, G., Lazarian, A., Vishniac, E.~T., \& Otmianowska-Mazur, K.\ 2009, \apj, 700, 63
\bibitem[Lazarian(2005)]{2005AIPC..784...42L} Lazarian, A.\ 2005, Magnetic 
Fields in the Universe: From Laboratory and Stars to Primordial 
Structures., 784, 42
\bibitem[Lazarian \& Vishniac(1999)]{lazarian99} Lazarian, A., Vishniac, E.~T. \ 1999, \apj, 512, 700, (LV99)
\bibitem[Lazarian \& Vishniac(2008)]{lazarian08} Lazarian, A., Vishniac, E.~T. \ 2008, arXiv:0812.2019
\bibitem{76} Lazarian, A., Vishniac, E., \& Cho, J.\ 2004, ApJ, 603, 180
\bibitem[Parker\ 1957]{P57}Parker, E.N.\ 1957, J. Geophys. Res., 62, 509
\bibitem{} Parker, E.N. 1979, Oxford, Clarendon Press; New York, Oxford University Press
\bibitem[Passot 
\& V{\'a}zquez-Semadeni(2003)]{passot2003} Passot, T., \& V{\'a}zquez-Semadeni, E.\ 2003, \aap, 398, 845
\bibitem[Petschek\ 1964]{P64} Petschek, H.E. 1964, 
{\it The Physics of Solar Flares}, AAS-NASA
Symposium, NASA SP-50 (ed. W.H. Hess), Greenbelt, Maryland, p.~425
\bibitem[]{} Santos-Lima, R., Lazarian, A., de Gouveia dal Pino, E. \& Cho, J. 2010, ApJ, submitted, arXiv 0910.1117
\bibitem[Shay et al.(1998)]{shay1998} Shay, M.~A., Drake, J.~F.,
Denton, R.~E., \& Biskamp, D.\ 1998, Journ. Geoph. Res., 103, 9165
\bibitem[Shu et al.(2007)]{2007ApJ...665..535S} Shu, F.~H., Galli, D., 
Lizano, S., Glassgold, A.~E., \& Diamond, P.~H.\ 2007, \apj, 665, 535
\bibitem[Sweet\ 1958]{S58} Sweet, P.A.\ 1958, in IAU Symp. 6, Electromagnetic 
Phenomena in 
Cosmical Plasma, ed. B. Lehnert (New York: Cambridge Univ. Press), 123
\bibitem[Tassis
\& Mouschovias(2005)]{tassis2005} Tassis, K., \& Mouschovias, T.~C.\ 2005, \apj, 618, 769
\bibitem[Uzdensky
\& Kulsrud(2000)]{uzdensky2000} Uzdensky, D.~A., \& Kulsrud, R.~M.\ 2000, Physics of Plasmas, 7, 4018
\bibitem[Yamada et al.(2006)]{yamada2006} Yamada, M., Ren, Y., Ji,
H., Breslau, J., Gerhardt, S., Kulsrud, R.,
\& Kuritsyn, A.\ 2006, Physics of Plasmas, 13, 052119
\end{thebibliography}
\end{document}